\documentclass[11pt]{article}

\usepackage{amsmath}
\usepackage{amssymb}
\usepackage{amsfonts}
\usepackage{amsthm}

\usepackage[utf8x]{inputenc} 

\usepackage{graphicx}
\usepackage[autostyle]{csquotes}
\usepackage[linktocpage]{hyperref}
\usepackage{fullpage}
\usepackage{slashed}
\usepackage{mathtools}
\usepackage{indentfirst}
\usepackage{bm}
\usepackage{mathrsfs} 
\usepackage{authblk}
\usepackage{tikz}
\usepackage{tikz-cd}

\usepackage{subfig}
\usepackage{makecell}
\usepackage{array}
\usepackage{float}

\DeclarePairedDelimiter\ceil{\lceil}{\rceil}
\DeclarePairedDelimiter\floor{\lfloor}{\rfloor}

\allowdisplaybreaks

\newtheorem{theorem}{Theorem}[section]

\newtheorem{conjecture}[theorem]{Conjecture}

\newcommand{\intg}{\mathbb{Z}}

\newcommand{\real}{\mathbb{R}}
\newcommand{\complex}{\mathbb{C}}

\newcommand{\rarw}{\rightarrow}

\newcommand{\lb}{\left(}
\newcommand{\rb}{\right)}
\newcommand{\lsb}{\left[}
\newcommand{\rsb}{\right]}
\newcommand{\lac}{\left\{}
\newcommand{\rac}{\right\}}

\newcommand{\RN}[1]{%
  \textup{\uppercase\expandafter{\romannumeral#1}}%
}

\title{Fiber sum formulas for 4-manifolds, topological modular forms and $6d\ \mathcal{N}=(1,0)$ theories}
\author{John Chae}
\affil{Altadena, California, USA \\ {yjchae@formerstudents.ucdavis.edu}}
\date{}  

\begin{document}

\maketitle

\begin{abstract}

Using the relation between four manifolds and topological modular form (TMF) from the six dimensional approach, we exhibit fiber sum formulas for infinite families of smooth spin four manifolds associated to compactifications of free and interacting $6d\ (1,0)$ SCFTs. We find that even the free theories have nontrivial fiber sum formulas and their forms are sensitive to an individual theory and parameters of four-manifolds. Furthermore, we reinforce the conjecture of Stolz and Teichner by expanding its evidence.\\

\end{abstract}

%\tableofcontents

\section{Introduction}

%\begin{enumerate}
	%\item Generalities 1: Existence of 6D SCFTs : superconformal algebra classification result, supermultitplet containing an energy-momentum tensor.
	%
	%\item Generalities 2: Constructions/Approaches: Supermultiplets, moduli spaces, anomaly free vs M-theory and F-theory.
	%
	%\item Connection to the TMF theory 
	%
	%\item The 6D SCFT approach and the invariants of smooth 4-manifolds $T{M^4]$
	%
	%
%\end{enumerate}

Conformal field theory (CFT) is a class of quantum field theory that possesses conformal symmetry and plays an important role in physics and mathematics. From physics viewpoint, CFTs describe phenomena in high energy systems and condensed matter systems. For the latter perspective, CFTs are intimately related to vertex operator algebra, its representation theory, and topology via topological quantum field theories.\\ 

Due to its multifaceted aspects, CFTs of various dimensions have been investigated extensively over the past decades. However, a class of CFTs endowed with supersymmetry, called superconformal field theory (SCFT), in six dimensions has been mysterious (see \cite{HR} for a review). Its enigmatic aspects originate from a 6-dimensional SCFT being strongly interacting, containing tensionless strings and having no known lagrangian. Consequently, the standard approaches to conventional (S)CFTs does not apply, which makes 6d SCFTs challenging to analyze. Furthermore, its existence is rather special. By the classification of superconformal algebra~\cite{Nahm}, the maximal spacetime dimension in which superconforomal algebra exists is six ($D \leq 6$). The maximal amount of supersymmetry that can be endowed to a 6d CFT is $\mathcal{N} \leq 2$ by an existence of a supermultiplet containing an energy-momentum tensor; this amounts to sixteen pseudo-real supercharges. This leads to two chiral $\mathcal{N}=(1,0)$ and $(2,0)$ SCFTs. Their superconforomal symmetry algebra is $osp(6,2 | \mathcal{N} )$ whose (maximal) bosonic subalgebra is $so(6,2) \times sp(\mathcal{N})_R$, where $R$ stands for R-symmetry. 
\newline

Despite 6-dimensional SCFTs are nonlagrangian theories, there are two approaches to analyze them. The first is string/M/F-theory and the second one is supermultiplets and moduli spaces. In the latter approach, one begins with a multiplet and explores corresponding moduli space of vacuum via RG flow and analyze anomaly phenomenon to extract information about the 6d SCFT. This strategy was used in \cite{CDI, I}. In case of a $6d\ (1,0)$ SCFT, there are three kinds of supermultiplets: hypermultiplet, vector multiplet and tensor multiplet. The first two are described in Section 4 and 5. The moduli space associated to hypermultiplet is called Higgs branch while tensor branch\footnote{A moduli space of vacuum does not exists for the vector multiplet since it does not contain a scalar field.} for tensor multiplet. The Higgs branch breaks $sp(\mathcal{N})_R$ symmetry whereas the tensor branch preserves it. All interacting $6d\ (1,0)$ theories have a tensor branch. 
\newline

Anomaly phenomenon is a core aspect of a QFT. It provides consistency conditions and serves as a tool for extracting nonperturative information of the theory. The latter part is often called 't Hooft anomaly matching, which is utilized, for example, in examining UV and IR theories connected by renormalization group flows.  In 6d SCFTs, four point correlation functions of currents associated symmetries represent anomalies. The symmetries can be flavor, gauge and/or diffeomorphsim. Roughly speaking, depending on whether the currents in the correlator are same or not, an anomaly is divided into pure and mixed types. For example, there are gauge anomaly and gravitational anomaly for the pure type and gauge-gravitational anomaly for a mixed type. Some anomalies whether pure or mixed must be canceled to obtain a consistent 6d SCFT\footnote{The correlation functions involving a gauge current must be canceled for any 6d SCFTs.}. A key object in the anomaly analysis is an anomaly polynomial of a theory. It encodes all perturbative anomaly information (see Appendix B for a review). As we will see, anomaly polynomials play a crucial role in this paper.
\newline     

A unique feature of 6d SCFTs is generating a variety of lower dimensional field theories through compactifications. An aspect of the theories relevant to our paper is a conjecture by Stolz and Teichner~\cite{ST1, ST2}. The conjecture predicts a connection between topological modular forms (TMF)~\cite{DFHH, H} and a space of 2-dimensional minimally suspersymmetric qunatum field theories. This conjecture was motivated by search for a geometric construction of an elliptic cohomology theory. A simplified\footnote{TMF is evaluated at a point in this paper, $\text{TMF}^{-d} (pt)=\text{TMF}_d (pt) = \pi_{d}\text{TMF}$. See \cite{ST2, GJF1} for the general version of the conjecture.} version proposes a classification of the 2d suspersymmetric theories via TMF, which is a generalized cohomology theory.
\begin{conjecture}(simplified\, \cite{ST2})
$$
\pi_{d} \text{TMF} \simeq \pi_0 \lac 2d	\ \mathcal{N}=(0,1)\ \text{SQFTs with gravitational anomaly}\ 2(c_R - c_L) = d \rac.
$$
\end{conjecture} 
\noindent On the RHS, using the fact that gravitational anomaly $d$ is a deformation invariant, it labels connected components of the space of the theories. Moreover, theories having the same value of the gravitational anomaly reside in the same connected component and they are connected by supersymmetry preserving continuous deformations, which include renormalization group flows. Namely,
$$
\text{RHS} = \lac 2d	\ \mathcal{N}=(0,1)\ \text{SQFTs with gravitational anomaly}\ 2(c_R - c_L) = d \rac / \text{deformations}
$$
Furthermore, the conjecture implies that geometric cocycles of the cohomology theory are represented by the field theories and predicts TMF-valued invariants for the $2d$ supersymmetric theories. The theory of topological modular forms have been playing a key role in various physics literature in the last few years~\cite{T,TY,LP, GJF2, GJFW}.
\newline

An investigation of Conjecture 1.1 in relation to an invariant of 4-manifolds was initiated in \cite{GPPV}. Specifically, evidence for the conjecture was provided for a range of homotopical degrees d (see Table 2 in \cite{GPPV}). Moreover, a connected sum formula of the TMF-valued invariant of 4-manifolds associated to compactifications of a \textit{nonphysical} $6d\ (1,0)$ SCFT on nonsmooth and nonspin 4-manifolds was proposed\footnote{A compactification of a \textit{physical} $6d\ (1,0)$ SCFT requires compact 4-manifolds to be smooth and spin.} (see Section 2.7 of \cite{GPPV}).\\

In this paper, we extend the investigation of \cite{GPPV} by analyzing compactifications of a \textit{physical} $6d\ (1,0)$ SCFT of both free and interacting theories on smooth, spin closed 4-manifolds. This class of 4-manifolds appear abundantly as surface bundles. We consider the fiber sum operation between these manifolds, which is more natural operation than the connected sum for the class of 4-manifolds~\cite{Gf} (see Appendix A for a review). We conjecture their fiber sum formulas for the above mentioned theories and provide evidence for the conjectures via the explicit examples summarized in tables. Furthermore, we strengthen Conjecture 1.1 by finding appropriate 2d theories corresponding to a broader range of homotopical degrees d than the ones considered in \cite{GPPV}. 
\newline

This paper is organized as follows. In Section 2, we expand the example list for the toy model as an exercise. We next describe allowed 4-manifolds in compactifications of (phyisical) $6d\ (1,0)$ SCFTs in Section 3. We begin with a free $6d\ (1,0)$ hypermultiplet theory compactified on  4-manifolds to find corresponding fiber sum formula in Section 4. And then we move onto a free $(1,0)$ vector multiplet theory in Section 5. We next consider an interacting theory, namely, the rank one E-string theory and obtain a fiber sum formula for specific 4-manifolds in Section 6. In Section 7, we reinforce Conjecture 1.1 by providing more evidence. We finish with open questions in Section 8.

%We note that when a theory on the RHS is realizable as a nonlinear sigma model, its target space needs to be compact.

%The original constructions of a 6d SCFT appeared in \cite{W,S} and then shown to be a conventional QFT in \cite{S}.\\ 

\section{The extended toy model}

A toy model was introduced in \cite{GPPV}. Since it is not associated with a $6d\ (1,0)$ SCFT, hence $2d\, (0,1)$ theories do not arise from a $6d$ SCFT. Furthermore, the model deals with a broader range of 4-manifolds compared to that of Section 3, namely,
$$
\frac{\lac \substack{ \text{Topological compact}\\  \text{oriented 4-manifolds} }\quad X   \rac}{\text{homeomorphism}} \xrightarrow{\makebox[1.5cm] T}  \frac{ \lac  \substack{ 2d\, \mathcal{N}=(0,1)\, \text{lattice SCFT} \\  \Gamma = H^2 (X ;\intg)/ Tor H^2 (X ;\intg) }  \rac }{\text{SUSY deformations}}  \xrightarrow{\makebox[1.5cm] S} \pi_{\ast} \text{TMF}
$$
\newline
\noindent We denote $\mathcal{T}:= S \circ T$ and hence $\mathcal{T}[X] \in \pi_{d}$ TMF, where the homotopical degree $d$ is given by $3 b_{2}^{+} - 2 b_{2}^{-}$ and $ \pi_{d}$ denotes the stable homotopy groups. 
\begin{table}[t]
\begin{center}
\begin{tabular}{ |c|c|c| } 
 \hline
$X$ & $d=3 b_{2}^{+} - 2 b_{2}^{-}$ & $\mathcal{T} [X] \in \pi_d \text{TMF}$  \\ 
 \hline
&&\\[-1em]
\makecell{ $ F_n $  \\ ($F_0 = S^2 \times S^2$) } & 1 & $\eta$ \\ 
 \hline
&&\\[-1em]
$ -E_8 $ & -16 & $E_4 / \Delta$ \\ 
 \hline
&&\\[-1em]
$\complex \mathbb{P}^2$ & 3 & $\nu$ \\ 
 \hline
&&\\[-1em]
$\overline{\complex \mathbb{P}^2}$ & -2 & 0 \\ 
 \hline
&&\\[-1em]
$K3$ & -29 & 0 \\ 
\hline
&&\\[-1em]
$\overline{K3}$ & 51 & $\nu \Delta^2$ \\ 
\hline
&&\\[-1em]
$\frac{1}{2} K3,\, E(1)$ & -15 & $\eta E_4 / \Delta$ \\ 
\hline
&&\\[-1em]
$\overline{\frac{1}{2} K3}$ & 25 & $\eta \Delta$ \\ 
\hline
\end{tabular}
\caption{This table is an expansion of the table in \cite{GPPV} by adding $-E_8 ,\, \overline{K3} ,\, \overline{\frac{1}{2} K3}$, and $E(1)$. $E(1)$ has the same degree as that of the Enriques surface $\frac{1}{2} K3$ but their intersection forms are inequivalent, hence they are not homeomorphic. $\eta$ and $\nu$ are generators of $\pi_1 \text{tmf}\simeq \intg_2$ and $\pi_3 \text{tmf}\simeq \intg_{24}$, respectively (see Appendix C for details).} 
\end{center}
\end{table}
\newline

We extend the examples of 4-manifolds in \cite{GPPV} for the toy model. The family of the Hirzebruch surfaces $F_n,\, n\geq 0$\footnote{$F_0 = S^2 \times S^2$ was done in  \cite{GPPV}}.
Its intersection form is
$$
Q_{F_n} = \begin{pmatrix}
n & 1 \\
1 & 0 
\end{pmatrix}
$$
where $n$ is the Euler number of the surface. Consequently, their $b_{2}^{\pm} (F_n) = 1$ so all $F_n$ have the same homotopical degree as $F_0$, which means that their $\pi_{d}$TMF coincide. Another well-known complex surface is a family of elliptic surfaces $E(n),\, n\in \intg_{+}$. This family includes $K3$ and its orientation reversal $\overline{K3}$. Using the data in Appendix A, we get $d = -14n-1 $ for $E(n)$.  Although $K3$-surface case is trivial, $\overline{K3}$ case doesn't vanish as shown in Table 1.  
%\newline

%
%\begin{align*}
%\sigma(Y_{32}) & = \left\langle L (p_1,\cdots , p_m) , [Y_{32}] \right\rangle\\
               %& \ni \frac{1}{488462349375} \lb 355554717 p_8 (Y_{32}) +  p_8\text{-independent terms} \rb \imp\\							
%p_8 (Y_{32})  &  = -85364982743040000, 							
%\end{align*}

%\clearpage

%\begin{table}[h!]
%\begin{center}
%\begin{tabular}{ |c|c|c|c| } 
 %\hline
%$X$ & $d  $  & $H(X)$  & $2D\ (0,1)$ theory generators  \\ 
 %\hline
%&&&\\[-0.5em]
%$E(2)$ & 4  & $\frac{2E_{4}^2 E_6}{\Delta}$ & $T_{(E_8 ;1)} \otimes \sigma (Y_8 \times Y_{12})$\\ 
 %\hline
%&&&\\[-0.2em]
%$ E(4) $ & 8 & $E_4  $    & $\sigma (SU(3)), \,  \sigma (Y_8)$ \\ 
 %\hline
%&&&\\[-0.2em]
%$E(6)$ & 12  & $ 2E_6 $   & $\sigma ( Y_{12})$ \\ 
 %\hline
%&&&\\[-0.2em]
%$E(20)$ & 40  &  $E_{4}^2 \Delta $       & $\sigma (Sp(2)^4), \quad \sigma (Y_8 \times Y_{32})$ \\ 
%\hline
%&&&\\[-0.2em]
%$E(22)$ & 44   & $2E_{4}^2 E_6$      & $ \sigma (Y_{12} \times Y_{32}) $ \\ 
%\hline
%&&&\\[-0.2em]
%$E(24)$ & 48  & 12\Delta^2 ,\ (j\Delta)^2     & $ \sigma ( Y_{48}), \, \sigma (Y_{8}^6 ) $ \\ 
%\hline
%&&\\[-0.5em]
%$\overline{E(2)}$ & -4 & $\frac{2E_{4} E_6}{\Delta} $ & \\ 
 %\hline
%&&\\[-0.5em]
%$\overline{E(4)}$ & -8 & $\frac{ E_{4}^2 }{\Delta} $  & \\ 
%\hline
%&&\\[-0.5em]
%$\overline{E(6)}$ & -12 & $\frac{2E_6}{\Delta} $ & \\ 
%\hline
%\end{tabular}
%\caption{}
%\end{center}
%\end{table}

\section{Four manifolds for compacification}

Before we analyze various physical theories, let us describe a class of 4-manifolds that can be used in compactification of a $6d\ (1,0)$ SCFT~\cite{GPPV}. In order to obtain a sensible $2d\ (0,1)$ theory from a compactification of a \textit{physical} $6d\ (1,0)$ theory, 4-manifolds need to be smooth and spin. In this setting, the diagram in Section 2 becomes~\cite{GPPV}
$$
\frac{\lac \substack{ \text{Smooth spin }\\  \text{compact 4-manifolds} }\quad X   \rac}{\text{diffeomorphism}} \xrightarrow{\makebox[1.5cm]{t}} \frac{ \lac  \substack{ 2d\, \mathcal{N}=(0,1)\, \text{theories} }  \rac }{\text{SUSY deformations}} \xrightarrow{\makebox[1.5cm]{$\sigma$}} \pi_{\ast} \text{TMF}
$$
where $\sigma$ is the toplogical Witten genus\footnote{Note there is a difference between the physical and the mathematical definitions of the map (see Appendix C). The former is slightly more general.}. Moreover, $H_{2}(X)$ being torsion-free is also assumed. This condition can be satisfied by choosing a simply connected $X$ ($\pi_1 (X) =1$). Furthermore, since the torsion-free part of $\pi_{\ast}$TMF is the elliptic genus and it is nonvanishing only when the degree $d$ of the TMF is $4\intg$, therefore we consider $X$ whose $d \in 4\intg$ in this paper.

\section{Hypermultiplet theory}

In this section, we consider a free $\mathcal{N}=(1,0)$ hypermultiplet theory in 6 dimensions. Although it is a free theory, a fiber sum formula for this theory turns out to be nontrivial (i.e. not a mere multiplication). This hypermultiplet consists of complex scalar fields $q$, $\tilde{q}^{\dagger}$ and their superpartner spinor fields $\psi$ on a spacetime manifold $M^{6}$; they are sections of certain bundles over $M^{6}$:
$$
\lac ( q, \tilde{q}^{\dagger} ) \in \Gamma \lb L  \otimes R \rb ,\quad  \psi \in \Gamma S \rac
$$
where $L$ is a trivial complex line bundle, $R$ is a principal $SU(2)_R$ bundle and $S$ is a spin bundle over $M^{6}$. $q$ and $\tilde{q}^{\dagger}$ together carry a fundamental representation of $SU(2)_R$ and their vacuum expectation values parametrizing the Higgs branch of the theory.
\newline

After compactification of the 6d free hypermultiplet theory on $X$, $2d\ (0,1)$ theory's field content depends the signature of $X$~\cite{GPPV}:
\begin{itemize}

	\item $\sigma > 0 : \frac{\sigma}{4}$ number of (0,1) Fermi multiplets
	
	\item $\sigma < 0 : \frac{| \sigma |}{4}$ number of (0,1) chiral multiplets
	
\end{itemize}

An advantage of 6-dimensional theories is that (perturbative) anomaly polynomials can be computed, which in turn can be used to obtain gravitational anomaly of compactifications of the theories (see Appendix B for review). In our context, we are interested in compatifications on smooth spin 4-manifolds, thus gravitational anomaly in 2 dimensions. The 2 dimensional gravitational anomaly associated to a free single $\mathcal{N}=(1,0) $ hypermultiplet is~\cite{GPPV}
$$
2(c_R - c_L) = -\frac{\sigma(X)}{4} \in \intg
$$
The above formula can be straightforwardly obtained from the anomaly polynomial of the theory (see Appendix B). An importance of the above formula is that the difference of the central charges is identified with the homotopical degree $d$ of $\pi_d \text{TMF}$ for any theories.\\

For the theory under consideration, its TMF degree\footnote{We use TMF degree and homotopical degree interchangeably (cf. footnote 3).} $d$ under the fiber sum is additive,
$$
d (X_1 \#_{f} X_2 ) = d(X_1) + d(X_2),
$$
since $\sigma$ is additive. We conjecture a (generalized) fiber sum formula for a $6d\ (1,0)$ free hypermultiplet theory. 
\begin{conjecture} 
Let $X_1 , X_2$ be smooth spin closed oriented 4-manifolds and $X$ be their (generalized) fiber sum $X= X_1 \#_f X_2 $, which is smooth spin closed and oriented. Moreover, let $d_1 , d_2$ be the TMF degrees of $X_1$ and $X_2$, respectively. For $d_1 + d_2 \notin 24\intg$, a $6d\ (1,0)$ free hypermultiplet SCFT compactified on $X_i , i=1,2$ satisfies the following (generalized) fiber sum formula for a generator of $\intg$-polynomial ring summand of $\pi_{\ast} \text{TMF}$ associated to $X$.
\begin{gather*} 
H( X_1 \#_{f} X_2 )  \circeq \Delta^{\floor*{ \frac{ w_1 + w_2 }{2} }} H(X_1) \ast H(X_2) \\
E_{4}^{p_1} \ast E_{4}^{p_2} = E_{4}^{p_1 + p_2},\quad \text{if}\ p_1 + p_2 \geq 3,\, \text{use}\ E_{4}^3 = j\Delta\\
E_{6}^{w_1} \ast E_{6}^{w_2}  = E_{6}^{b},\quad w_1 + w_2 \equiv b\quad \text{mod}\, 2, \quad  (b=0,1)\\
\Delta^{m_1} \ast \Delta^{m_2}  = \Delta^{m_1 + m_2},
\end{gather*}
\end{conjecture}
\noindent where the circle denotes up to the $j$-modular function.\\

\noindent\textbf{Remark 1}\ As stated in the conjecture, the (entire) RHS is understood as extracting only generators of the polynomial ring.\\
\noindent\textbf{Remark 2}\ Since we consider the total degree $\notin 24\intg$, the $j$-function is discarded in a final expression on the RHS.\\

In the next subsections, we provide evidence for this conjecture.

%For the class of $X$ of our interest, $\sigma(X) \in 16\intg$. This value implies that~\cite{DFHH}
%$$
%\pi_{-\frac{\sigma}{4}} \text{TMF}  \supseteq \intg[x].
%$$

%\clearpage

\begin{table}[t!]
\begin{center}
\begin{tabular}{ |c|c|c| } 
 \hline
$X$ & $d  $  & $H(X)$   \\ 
 \hline
&&\\[-0.5em]
$E(2)$ & 4  & $\frac{2E_{4}^2 E_6}{\Delta}$ \\ 
 \hline
&&\\[-0.5em]
$ E(4) $ & 8 & $E_4  $    \\ 
 \hline
&&\\[-0.5em]
$E(6)$ & 12  & $ 2E_6 $    \\ 
 \hline
&&\\[-0.5em]
$E(20)$ & 40  &  $E_{4}^2 \Delta $       \\ 
\hline
&&\\[-0.5em]
$E(22)$ & 44   & $2E_{4} E_6 \Delta $      \\ 
\hline
&&\\[-0.5em]
$E(24)$ & 48  & $12\Delta^2 ,\ (j\Delta)^2$    \\ 
\hline
&&\\[-0.5em]
$\overline{E(2)}$ & -4 & $\frac{2E_{4} E_6}{\Delta} $ \\ 
 \hline
&&\\[-0.5em]
$\overline{E(4)}$ & -8 & $\frac{ E_{4}^2 }{\Delta} $  \\ 
\hline
&&\\[-0.5em]
$\overline{E(6)}$ & -12 & $\frac{2E_6}{\Delta} $  \\ 
\hline
\end{tabular}
\caption{Generators $H(X)$ of $\intg$-polynomial ring part of $\pi_{d}$TMF associated to $E(2r)$ and its orientation reversal $\overline{E(2r)}$  for the free $(1,0)$ hypermultiplet theory. The third column can be computed using \cite{DFHH} (see Chapter 13) and Appendix C.}
\end{center}
\end{table}

\subsection{Elliptic surfaces}

We first consider a family of smooth complex elliptic surfaces $E(n)$. They can be obtained inductively beginning from $E(1)$ and a fiber sum operation (see Appendix A). For even values of $n$, they are spin manifolds. Using their topological data in Appendix A, homotopical degree $d$ of $E(n)$ is
$$
d= -\frac{\sigma}{4} =  4r,\quad r= 1,2,3,\cdots
$$
We list generators of integer polynomial ring part of $\pi_{d}\text{TMF}$ in Table 2. Conjecture 4.1 was checked, for example, using the table.\\

The generators in Table 2 can be computed using \cite{DFHH} (see Chapter 13) and Appendix C. For example, $E(20)$ has $\sigma= -160$ and hence homotopical degree $d=40 \equiv 0$ mod 8. This implies that $\pi_{40}$TMF contains a $\intg[x]$ (see Appendix C). From \cite{DFHH}, we see that prime 2 contribution of torsion free part of $\pi_{40}\text{tmf}_{(2)}$ is  $\intg_{(2)}^{2}$ whereas prime 3 contribution of $\pi_{40}\text{tmf}_{(3)}$  is $\intg_{(3)}^{2}$\footnote{Torsion from prime 2 is $\intg / 4\intg$ whereas $\intg / 3\intg$ from prime 3.}. They are both generated by $E_{4}^{2}\Delta$. After localization, we arrive at the polynomial ring with $\intg$ coefficient for torsion free part of $\pi_{40}$TMF.
\newline

\noindent For negative values of $d$ such as $d=-8 \equiv 0$ mod 8, we need to use the 576 periodic property of $\pi_{d}$TMF. Namely,
$$
\pi_{-8}\text{TMF}  \simeq \pi_{568}\text{TMF}. 
$$
We next use the fact that $\pi_{\ast}\text{tmf}_{(2)}$ is 192 periodic and $\pi_{\ast}\text{tmf}_{(3)}$ is 72 periodic. Hence, we easily deduce that $\pi_{184}\text{tmf}_{(2)}$ for the former and $\pi_{64}\text{tmf}_{(3)}$ for the latter. Their torsion free parts are generated by $E_{4}^2 \Delta^7$ and $E_{4}^2 \Delta^2$, respectively~\cite{DFHH}. After localization, we arrive at $E_{4}^2 / \Delta$ for $\pi_{-8}$TMF.
\newline

%We note that the logarithmic transformation of $E(2r)$ preserves their $\pi_d \text{TMF}$ since it doesn't change the Betti numbers of $E(2r)$. Therefore, $\pi_d \text{TMF}$ cannot detect the log. transforms on $E(2r)$.
%\newline
%\pagebreak

\subsection{Fiber sum manifolds}

\begin{table}[t]
\begin{center}
\begin{tabular}{ |c|c|c| } 
 \hline
$X$ & $d  $ & $ H(X) $    \\ 
 \hline
&&\\[-0.2em]
\makecell{ $Z^{3}_{2,3} = X_{2,3} \#_{f} E(6)_K$ \\ $V^{3}_{3} = X_{3} \#_{f} E(6)_K$ } & -4 &  $\frac{2E_4 E_6}{\Delta}$    \\
 \hline
&&\\[-0.2em]
\makecell{ $Z^{2}_{2,3} = X_{2,3} \#_{f} E(4)_K$ \\ $V^{2}_{3} = X_{3} \#_{f} E(4)_K $ } & -8 &  $\frac{ E_{4}^2 }{\Delta}$  \\ 
 \hline
&&\\[-0.2em]
\makecell{ $ Z^{1}_{2,3} = X_{2,3} \#_{f} E(2)_K $ \\ $V^{1}_{3} = X_{3} \#_{f} E(2)_K $ }  & -12 & $\frac{2E_{6} }{\Delta}$   \\ 
\hline
&&\\[-0.2em]
 $V^{16}_{5} = X_{5} \#_{f} E(32)_K $   & -16 &  $\frac{ E_{4} }{\Delta}$   \\
\hline
&&\\[-0.2em]
$Z^{12}_{2,5} = X_{2,5} \#_{f} E(24)_K $ & -32 &     $\frac{ E_{4}^2 }{\Delta^2}$    \\ 
\hline
&&\\[-0.2em]
$Z^{11}_{2,5} = X_{2,5} \#_{f} E(22)_K $ & -36 &   $\frac{ 2E_{6} }{\Delta^2}$     \\ 
\hline
&&\\[-0.2em]
\makecell{ $Z^{10}_{2,5} = X_{2,5} \#_{f} E(20)_K $ \\ $ V^{10}_{5} = X_{5} \#_{f} E(20)_K$ }  & -40 &    $\frac{ E_{4} }{\Delta^2}$     \\ 
\hline
\end{tabular}
\caption{Generators $H(X)$ of $\intg$-polynomial ring part of $\pi_{d}$TMF associated to $Z^{r}_{2,n}$ and $V^{r}_{n}$ for the free $(1,0)$ hypermultiplet theory are listed. For $ X_{2,3}$ and $X_{3}$, their TMFs are $\pi_{-16} \text{TMF} = \left\langle E_{4}/\Delta \right\rangle$. For $X_{2,5}$ and $X_{5}$, their TMFs are $\pi_{-80} \text{TMF} = \left\langle E_{4}^2/ \Delta^4 \right\rangle$. $E(2r)_K$ has the same the Euler characteristic and the signature as $E(2r)$.}
\end{center}
\end{table}

For verifications of Conjecture 4.1, we analyze families of smooth spin irreducible 4-manifolds $Z^{r}_{2,n}$ and $V^{r}_{n}$ constructed in \cite{AP}. Their constructions and properties are summarized in Appendix A. The homotopical degrees of $Z^{r}_{2,n}$ and $V^{r}_{n}$ coincide
$$
d= -\frac{\sigma}{4} = -\frac{2 n^3}{3}+\frac{2 n}{3}+4 r.
$$
Their generators $H(X)$ of polynomial rings of $\pi_{\ast}$TMF are listed in Table 3. Using this table and Table 2, we verified Conjecture 4.1.\\

We note that $X_{2,3}$ factor of $Z^{r}_{2,3}$ is $\Sigma_6$-bundle over $\Sigma_{19}$ whereas $X_{2,5}$ in $Z^{r}_{2,5}$ is $\Sigma_{10}$-bundle over $\Sigma_{51}$. $X_3$ in $V^{r}_{3}$ is $\Sigma_{9}$-bundle over $\Sigma_{19}$; $X_5$ in $V^{r}_{5}$ is $\Sigma_{15}$-bundle over $\Sigma_{51}$. 
\newline
%$$
%X_{2,3},\, X_{3} : \pi_{-16} \text{TMF} =  \intg[x] \left\langle \frac{E_{4}}{\Delta} \right\rangle \qquad X_{2,5} ,\, X_{5} : \pi_{-80} \text{TMF} =  \intg[x] \left\langle \frac{E_{4}^2}{\Delta^4} \right\rangle
%$$
%\noindent We tabulated a sample of evidence for the conjecture in Table 3. 

%\noindent\textbf{Remark 3}\  In case $d_1 + d_2 = 24m,\, m \in Z^{\ast}$, we may obtain $\Delta$ part of a (full) generator $( 24/gcd(24, m) ) \Delta^{m}$ of an infinite cyclic group summand of $\pi_d TMF$.\\

\begin{table}
\begin{center}
\begin{tabular}{ |c|c|c| } 
 \hline
$X$ & $ d $ & $ H(X)$  \\ 
 \hline
&&\\[-0.2em]
$Z(1,1)_n $ & $ 8 $ &  $E_4$ \\
 \hline
&&\\[-0.2em]
$ Z(1,3)_n $ & $ 12$ & $2E_6$ \\
 \hline
&&\\[-0.2em]
$ Z(3,3)_n $ & $ 16 $ &  $E_{4}^{2}$ \\
 \hline
&&\\[-0.2em]
$ Z(1,7)_n $ & $ 20$  & $2E_4 E_6$ \\
 \hline
%&&\\[-0.2em]
%$ Z(5,5)_n $ & $ \pi_{-28} \text{TMF} \simeq  \intg / 2\intg \left\langle \frac{ \kappa \nu^2 }{\Delta^2} \right\rangle \oplus  \intg[x] \left\langle \frac{ 2 E_{4} E_{6} }{\Delta^2} \right\rangle  $ & \\
 %\hline
&&\\[-0.2em]
$ Z(3,11)_n $ & $ 32 $ & $E_4 \Delta$  \\
 \hline
%&&\\[-0.2em]
%$ Z(5,11)_n $ &  $ 36 $ & $2E_6 \Delta $\\
 %\hline
\end{tabular}
\caption{Generators $H(X)$ of $\intg$-polynomial ring part of $\pi_{d}$TMF associated to $ Z(k,m)_n$ for the free $(1,0)$ hypermultiplet theory are listed.}
\end{center}
\end{table}

\subsection{Nonsymplectic manifolds}

We provide further evidence of Conjecture 4.1. Specifically, we  consider a family $Z(k,m)_n $ of manifolds that are smooth spin nonsymplectic 4-manifolds constructed in \cite{Sz} (see Appendix A for details). Their homotopical degree is
$$
d= -\frac{\sigma}{4} = 2 (k+m) + 4
$$ 
for all $k,m \geq 1$ and odd. Conjecture 4.1 was checked using Tables 4 and 2, namely, $Z(k,m)_n \#_f Z(r,s)_n $ and $Z(k,m)_n \#_f E(2r) $ and $Z(k,m)_n \#_f \overline{E(2r)} $.
%$$
%V^{1}_{3} = X_{3} \#_{f} E(2)_K
%$$
%$$
%V^{1}_{3} : \pi_{-12}  \text{TMF} =  \intg[x] \left\langle \frac{ 2 E_{6}  }{\Delta} \right\rangle
%$$
%
%$$
%V^{2}_{3} = X_{3} \#_{f} E(4)_K
%$$
%$$
%V^{2}_{3} : \pi_{-8}  \text{TMF} =  \intg[x] \left\langle \frac{  E_{4}^2  }{\Delta} \right\rangle
%$$

%$$
%V^{3}_{3} = X_{3} \#_{f} E(6)_K
%$$
%$$
%V^{3}_{3} : \pi_{-4}  \text{TMF} =  \intg[x] \left\langle \frac{  2E_{4}E_{6}  }{\Delta} \right\rangle
%$$

%$$
%V^{10}_{5} = X_{5} \#_{f} E(20)_K
%$$
%$$
%V^{10}_{5} : \pi_{-40}  \text{TMF} =  \intg[x] \left\langle \frac{  E_{4} }{\Delta^2} \right\rangle
%$$

%$$
%V^{16}_{5} = X_{5} \#_{f} E(32)_K
%$$
%$$
%V^{16}_{5} : \pi_{-16}  \text{TMF} =  \intg[x] \left\langle \frac{  E_{4} }{\Delta} \right\rangle
%$$

\section{Vector multiplet theory}

In this section, we describe a free $\mathcal{N}=(1,0)$ vector multiplet theory in 6 dimensions. We note that the theory is only scale invariant (nonconformal)~\cite{CDI,SNR}. A fiber sum formula depends on a value of parameter $n$. This multiplet consists of a 1-form $A$ and its superpartner spinor fields $\bar{\lambda}$ on $M^{6}$:
$$
\lac A \in \Omega^{1}(M^{6}),\quad  \bar{\lambda} \in \Gamma \lb S \otimes R \rb \rac,
$$
where $S$ is a spin bundle, $R$ is a principal $SU(2)_R$ bundle over $M^{6}$ and $\bar{\lambda}$ carries the fundamental representation of $SU(2)_R$. 
\newline

After compactification of the 6d free hypermultiplet theory on $X$, $2d\ (0,1)$ theory's field content depends the Betti numbers $b_i $ of $X$~\cite{GPPV}:
\begin{itemize}

	\item $ b_0$ number of (0,1) vector multiplets
	
	\item $ b_1$ number of (0,1) compact chiral multiplets
	
	\item $ b_{2}^{-}$ number of (0,1) Fermi multiplets
	
\end{itemize}
%\newline
The gravitational anomaly of a $2d\ (0,1)$ theory obtained from a compactification of the 6D free $\mathcal{N}=(1,0) $ vector multiplet theory is~\cite{GPPV}
$$
d = 2(c_R - c_L) = - 2\chi_h (X) \in \intg
$$
where $\chi_h$ is the holomorphic Euler characteristic of $X$. This formula is obtained from the anomaly polynomial of the theory (see Appendix B)~\footnote{The anomaly polynomial takes into account the nonconformality of the theory.}.
\newline

For this theory, the TMF degree $d$ under the fiber sum receives a contribution from the gluing surface $\Sigma$,
$$
d (X_1 \#_{f} X_2 ) = d(X_1) + d(X_2) + \chi (\Sigma).
$$
When $\Sigma = T^2$, then it is additive.

\subsection{Elliptic surfaces}

We begin with the elliptic surface, whose results are useful in the next subsection. The TMF degree for $E(2r)$ is $-\chi_h = -4r$, which is same as that of $\overline{E(2r)}$ of the free hypermultiplet in the previous section. For $\overline{E(2r)}$, its degree is $-\chi_h = -20r$ so the TMF of the orientation reversed manifold is a subset of $E(2r)$'s. Hence, the vector multiplet is insensitive to the orientation reversal.
\newline
As an example, we choose $E(26)$ in Table 5. From the periodic property of $\pi_{\ast}$TMF, $\pi_{-52} \text{TMF} \cong \pi_{524} \text{TMF}$. Using \cite{DFHH},
$$
\pi_{140}\text{tmf}_{(2)} = \intg_{(2)}^6 \qquad \pi_{20}\text{tmf}_{(3)} = \intg/3\intg \oplus \intg_{(3)}
$$
The torsion free parts are generated by $2E_4 E_6 \Delta^5$ and  $E_4 E_6$, respectively. After localization and using that $d(\Delta)=24$, we get the generator in the table.

\begin{table}[t!]
\begin{center}
\begin{tabular}{ |c|c|c| } 
 \hline
$X$ & $ d  $ & $ V(X) $   \\ 
 \hline
&&\\[-0.5em]
\makecell{ $E(4)$ } & -8 & $  \frac{ E_{4}^2  }{\Delta} $   \\ 
 \hline
&&\\[-0.5em]
\makecell{ $E(6)$ } & -12 & $  \frac{2 E_{6} }{\Delta} $   \\ 
 \hline
&&\\[-0.5em]
\makecell{ $E(8)$ } & -16 & $  \frac{E_{4} }{\Delta} $   \\ 
 \hline
&&\\[-0.5em]
\makecell{ $E(20)$  } & -40 & $ \frac{ E_{4} }{\Delta^{2}} $  \\
 \hline
&&\\[-0.5em]
\makecell{$E(24)$  } & -48 & $ \frac{ E_{4}^3 }{\Delta^{3}} $  \\
 \hline
&&\\[-0.5em]
\makecell{$E(26)$ }  & -52 & $ \frac{ 2E_{4} E_{6}  }{\Delta^{3}} $  \\ 
\hline
&&\\[-0.5em]
\makecell{ $E(28)$ } & -56 & $   \frac{ E_{4}^2 }{\Delta^{3}} $  \\ 
\hline
\end{tabular}
\qquad
\begin{tabular}{ |c|c|c| }
 \hline
$X$ & $ d  $ & $ V(X) $   \\ 
 \hline
%&&\\[-0.5em]
%\makecell{ $ Z^{1}_{2,3} = X_{2,3} \#_{f} E(2)_K $  }  & -264 & $ 24\intg \left\langle \frac{24}{\Delta^{11}} \right\rangle + x\intg [x]  \left\langle \frac{E_{4}^3}{\Delta^{12}} \right\rangle  $   \\ 
%\hline
&&\\[-0.5em]
\makecell{ $Z^{2}_{2,3} = X_{2,3} \#_{f} E(4)_K$} & -268 & $  \frac{2E_{4}E_{6} }{\Delta^{12}} $   \\ 
 \hline
&&\\[-0.5em]
\makecell{ $Z^{3}_{2,3} = X_{2,3} \#_{f} E(6)_K$  } & -272 & $ \frac{ E_{4}^2 }{\Delta^{12}} $  \\
 \hline
&&\\[-0.5em]
\makecell{ $Z^{4}_{2,3} = X_{2,3} \#_{f} E(8)_K$  } & -276 & $ \frac{ 2E_{6} }{\Delta^{12}} $  \\
 \hline
&&\\[-0.5em]
\makecell{ $Z^{10}_{2,5} = X_{2,5} \#_{f} E(20)_K $ }  & -1220 & $ \frac{ 2E_{4}^2 E_{6}  }{\Delta^{52}} $  \\ 
\hline
%&&\\[-0.5em]
%\makecell{ $Z^{11}_{2,5} = X_{2,5} \#_{f} E(22)_K $ } & -1224 & $ 8 \intg \left\langle \frac{ 8 }{\Delta^{51}} \right\rangle + x\intg[x] \left\langle ? \right\rangle $  \\ 
%\hline
&&\\[-0.5em]
\makecell{ $Z^{12}_{2,5} = X_{2,5} \#_{f} E(24)_K $ } & -1228 & $   \frac{2 E_{4} E_{6} }{\Delta^{52}} $  \\ 
\hline
&&\\[-0.5em]
\makecell{ $Z^{13}_{2,5} = X_{2,5} \#_{f} E(26)_K $ } & -1232 & $   \frac{ E_{4}^2 }{\Delta^{52}} $  \\ 
\hline
%&&\\[-0.5em]
%\makecell{ $Z^{14}_{2,5} = X_{2,5} \#_{f} E(28)_K $ } & -1236 & $   \frac{ 2E_{6} }{\Delta^{52}} $  \\ 
%\hline
&&\\[-0.5em]
\makecell{ $Z^{15}_{2,5} = X_{2,5} \#_{f} E(28)_K $ } & -1240 & $   \frac{ E_{4} }{\Delta^{52}} $  \\ 
\hline
\end{tabular}
\end{center}
\caption{Generators $V(X)$ of $\intg$-polynomial ring part of $\pi_{d}$TMF associated to $Z^{r}_{2,n}$ for the free $(1,0)$ vector multiplet theory are listed. The third column can be computed using \cite{DFHH} (see Chapter 13) and Appendix C.}
\end{table}

\subsection{Fiber sum manifolds}

\begin{table}[t!]
\begin{center}
\begin{tabular}{ |c|c|c| }
 \hline
$X$ & $ d  $ & $ V(X) $   \\ 
 \hline
&&\\[-0.5em]
\makecell{ $Z^{1}_{2,7} = X_{2,7} \#_{f} E(2)_K$} & -3224 & $  \frac{E_{4}^2}{\Delta^{135}} $   \\ 
 \hline
&&\\[-0.5em]
\makecell{ $Z^{2}_{2,7} = X_{2,7} \#_{f} E(4)_K$ } & -3228 & $ \frac{ 2E_{6} }{\Delta^{135}} $  \\
 \hline
&&\\[-0.5em]
\makecell{  $Z^{3}_{2,7} = X_{2,7} \#_{f} E(6)_K$} & -3232 & $ \frac{ E_{4} }{\Delta^{135}} $  \\
 \hline
&&\\[-0.5em]
\makecell{  $Z^{4}_{2,7} = X_{2,7} \#_{f} E(8)_K$}  & -3236 & $ \frac{ 2E_{4}^2 E_{6} }{\Delta^{136}} $  \\ 
\hline
&&\\[-0.5em]
\makecell{  $Z^{6}_{2,7} = X_{2,7} \#_{f} E(12)_K$}  & -3244 & $ \frac{ 2E_{4}E_{6} }{\Delta^{136}} $  \\ 
\hline
&&\\[-0.5em]
\makecell{  $Z^{7}_{2,7} = X_{2,7} \#_{f} E(14)_K$}  & -3248 & $ \frac{ E_{4}^2 }{\Delta^{136}} $  \\ 
\hline
\end{tabular}
\end{center}
\caption{Generators $V(X)$ of $\intg$-polynomial ring part of $\pi_{d}$TMF associated to $Z^{r}_{2,7}$ for the free $(1,0)$ vector multiplet theory are listed. The third column can be computed using \cite{DFHH} (see Chapter 13) and Appendix C.}
\end{table}

We consider $Z^{r}_{2,n}, n=3,5$ and $Z^{r}_{2,7}$ (see Appendix A for a review) and observe the following behaviors under the fiber sum.\\
\begin{conjecture} Let $X_1 = X_{2,n} , X_2 = E(2r)_K$ be smooth spin closed oriented 4-manifolds and $Z^{r}_{2,n}$ be their (generalized) fiber sum $ X_1  \#_{f} X_2 $, which is also smooth, spin, closed and oriented. Moreover, let $d_1 , d_2$ be the TMF degrees of $X_1 , X_2 $, respectively.  For a $6d\ (1,0)$ free vector multiplet SFT compactified on $X_i,\, i=1,2$ a fiber sum formula for $ d_1\notin 24\intg$, $d(Z^{r}_{2,n}) \notin 24\intg$ and $n=3,5$, is
\begin{gather*}
V(Z^{r}_{2,n}) \circeq \frac{1}{\Delta^s} V(X_{2,n}) \ast V(E(2r)_K)\quad \text{for all}\ r, \\
E_{4}^{p_1} \ast E_{4}^{p_2} = E_{4}^{p_1 + p_2},\quad \text{if}\ p_1 + p_2 \geq 3,\, \text{use}\ E_{4}^3 = j\Delta\\
E_{6}^{w_1} \ast E_{6}^{w_2}  = E_{6}^{b},\quad w_1 + w_2 \equiv b\quad \text{mod}\, 2, \quad  (b=0,1)\\
\Delta^{m_1} \ast \Delta^{m_2}  = \Delta^{m_1 + m_2}\\
s= \begin{cases}
n - \ceil*{\frac{p_1 + p_2}{n-1}} + (w_1 + w_2) ,\ r=\ \text{even} 
\\[10pt]
n - \ceil*{\frac{p_1 + p_2}{n-2}} ,\ r=\ \text{odd}\\
\end{cases}
\end{gather*}\\
Furthermore, in case of $Z^{r}_{2,7}$, for $ d_1 (X_{2,7}) \notin 24\intg$ and  d($Z^{r}_{2,7}$) $ \notin 24\intg$, we have
\begin{gather*}
V(Z^{r}_{2,7}) \circeq \frac{1}{\Delta^s} V(X_{2,7}) \ast V(E(2r)_K)\quad \text{for all}\ r, \\
E_{4}^{p_1} \ast E_{4}^{p_2} = E_{4}^{p_1 + p_2 -1},\quad \text{if}\ p_1 + p_2 -1 \geq 3,\, \text{use}\ E_{4}^3 = j\Delta\\
E_{6}^{w_1} \ast E_{6}^{w_2}  = E_{6}^{b},\quad w_1 + w_2 \equiv b\quad \text{mod}\, 2, \quad  (b=0,1)\\
\Delta^{m_1} \ast \Delta^{m_2}  = \Delta^{m_1 + m_2}\\
s= \begin{cases}
8,\ r=\ \text{odd} \\
9,\ r=\ \text{even}\\
\end{cases}
\end{gather*}
\end{conjecture}
\noindent \textbf{Remark}\, It can be inferred that the degree of $Z^{r}_{2,n}$ is $24\intg$, if the RHS consists of solely $\Delta$.\\

\noindent Verifications of Conjecture 5.1 were done for instance using Tables 5 and 6.

\section{E-strings}

We analyzed free theories in the previous sections, we proceed to examine an interacting $6D\ (1,0)$ SCFT, E-string theory compactified on a 4-manifold. We give a brief review of the theory and then move onto their compactifications and fiber sum formula.\\

\noindent \underline{Review}\ In M-theory, E-string theory of rank $1$ is a world volume theory of a brane system consisting of a M5-brane attached to a M9-brane, which carries an $E_8$ gauge symmetry. Specifically, the M5 is embedded as an E8-instanton. The low energy limit yields a single free hypermultiplet characterizing the center of mass motion of the M5-brane and an interacting $6d\ (1,0)$ superconformal field theory with an $E_8$ flavor symmetry. Furthermore, the strings arising from M2-branes suspended between the M5-brane and the M9-brane become massless (i.e. tensionless) as the M5-brane approaches the M9-brane. These strings are called E-strings. Anomaly polynomial of this theory was found in \cite{OST, OSTY} and recorded in Appendix B.\\

\begin{table}[t!]
\begin{center}
\begin{tabular}{ |c|c|c| } 
 \hline
$X$ & $ d  $ & $ E(X) $   \\ 
 \hline
&&\\[-0.5em]
\makecell{ $E(2)$ } & 116 & $  2E_4 E_6 \Delta^4 $   \\ 
\hline
&&\\[-0.5em]
\makecell{ $E(4)$ } & 232 & $  E_{4}^2 \Delta^9 $   \\ 
 \hline
&&\\[-0.5em]
\makecell{ $E(6)$ } & 348 & $ 2E_6 \Delta^{14} $   \\ 
 \hline
&&\\[-0.5em]
\makecell{ $E(8)$ } & 464 & $ E_{4} \Delta^{19} $   \\ 
 \hline
&&\\[-0.5em]
\makecell{ $E(10)$ } & 580 & $ 2E_{4}^2 E_{6} \Delta^{23} $   \\ 
 \hline
%&&\\[-0.5em]
%\makecell{ $E(20)$  } & -40 & $ \frac{ E_{4} }{\Delta^{2}} $  \\
 %\hline
%&&\\[-0.5em]
%\makecell{$E(24)$  } & -48 & $ \frac{ E_{4}^3 }{\Delta^{3}} $  \\
 %\hline
%&&\\[-0.5em]
%\makecell{$E(26)$ }  & -52 & $ \frac{ 2E_{4} E_{6}  }{\Delta^{3}} $  \\ 
%\hline
%&&\\[-0.5em]
%\makecell{ $E(28)$ } & -56 & $   \frac{ E_{4}^2 }{\Delta^{3}} $  \\ 
%\hline
\end{tabular}
\end{center}
\caption{Generators $E(X)$ of $\intg$-polynomial ring part of $\pi_{d}$TMF associated to $E(2r)$ for the E-string theory are listed. The third column can be computed using \cite{DFHH} (see Chapter 13) and Appendix C.}
\end{table}

\subsection{Elliptic surfaces}

From the anomaly polynomial of the E-strings and the topological data of $E(2r)_K$ in the appendices, we find its homotopical degree $d$ to be
$$
d = -2 ( 5 \sigma + 11 \chi_h) =  116r.
$$
Furthermore, using the fiber sum definition of $E(n)$ (see Appendix A), we easily deduce $E(m) \#_{f} E(n) = E(m+n) $.\\
\begin{conjecture} Let $E(2m)$ be an elliptic surface. For the E-string theory compactified on $E(2m)$, a fiber sum formula $E(2r) \#_{f} E(2s)$ in case of total homotopical degree $d=d(E(2r))+ d( E(2s)) \notin 24\intg$ is
\begin{gather*}
E( E(2r) \#_{f} E(2s))  \circeq  \Delta^{ \floor{\frac{w_1 + w_2 } {2} } }    E( E(2r) ) \ast E( E(2s) )\\
E_{4}^{p_1} \ast E_{4}^{p_2} = E_{4}^{p_1 + p_2 },\quad \text{if}\ p_1 + p_2 \geq 3,\, \text{use}\ E_{4}^3 = j\Delta\\
E_{6}^{w_1} \ast E_{6}^{w_2}  = E_{6}^{b},\quad w_1 + w_2 \equiv b\quad \text{mod}\, 2, \quad  (b=0,1)\\
\Delta^{m_1} \ast \Delta^{m_2}  = \Delta^{m_1 + m_2}\\
\end{gather*}
\end{conjecture}
\noindent As before, the circle denotes up to the j-modular function. This result coincides with that of the free $(1,0)$ hypermultiplet theory in Section 4. This is because the generic fiber of $E(n)$ is $T^2$ whose Euler characteristic vanishes and hence $d$ becomes additive for this manifold. It is straightforward to check the above conjecture via Table 7. 
\begin{table}[t!]
\begin{center}
\begin{tabular}{ |c|c|c| } 
 \hline
$X$ & $ d  $ & $E(X) $   \\ 
 \hline
%&&\\[-0.5em]
%\makecell{ $ Z^{1}_{2,3} = X_{2,3} \#_{f} E(2)_K $  }  & -264 & $ 24\intg \left\langle \frac{24}{\Delta^{11}} \right\rangle + x\intg [x]  \left\langle \frac{E_{4}^3}{\Delta^{12}} \right\rangle  $   \\ 
%\hline
&&\\[-0.5em]
\makecell{ $Z^{2}_{2,3} $} & -3268 & $  \frac{2E_{4}E_{6} }{\Delta^{137}} $   \\ 
 \hline
&&\\[-0.5em]
\makecell{ $Z^{3}_{2,3} $  } & -3152 & $ \frac{ E_{4}^2 }{\Delta^{132}} $  \\
 \hline
&&\\[-0.5em]
\makecell{ $Z^{4}_{2,3}$  } & -3036 & $ \frac{ 2E_{6} }{\Delta^{127}} $  \\
 \hline
&&\\[-0.5em]
\makecell{ $Z^{1}_{2,5}  $ }  & -16064 & $ \frac{ E_{4}^2  }{\Delta^{670}} $  \\ 
\hline
&&\\[-0.5em]
\makecell{ $Z^{2}_{2,5}  $ }  & -15948 & $ \frac{ 2E_{6}  }{\Delta^{665}} $  \\ 
\hline
&&\\[-0.5em]
\makecell{ $Z^{3}_{2,5}  $ }  & -15832 & $ \frac{ E_{4}  }{\Delta^{660}} $  \\ 
\hline
%&&\\[-0.5em]
%\makecell{ $Z^{4}_{2,5}  $ } & -15716 & $   \frac{2 E_{4}^2 E_{6} }{\Delta^{656}} $  \\ 
%\hline
%&&\\[-0.5em]
%\makecell{ $Z^{13}_{2,5}  $ } & -1232 & $   \frac{ E_{4}^2 }{\Delta^{52}} $  \\ 
%\hline
%&&\\[-0.5em]
%\makecell{ $Z^{14}_{2,5} = X_{2,5} \#_{f} E(28)_K $ } & -1236 & $   \frac{ 2E_{6} }{\Delta^{52}} $  \\ 
%\hline
%&&\\[-0.5em]
%\makecell{ $Z^{15}_{2,5} $ } & -1240 & $   \frac{ E_{4} }{\Delta^{52}} $  \\ 
%\hline
\end{tabular}
\qquad
\begin{tabular}{ |c|c|c| } 
 \hline
$X$ & $ d  $ & $E(X) $   \\ 
 \hline
%&&\\[-0.5em]
%\makecell{ $ Z^{1}_{2,3} = X_{2,3} \#_{f} E(2)_K $  }  & -264 & $ 24\intg \left\langle \frac{24}{\Delta^{11}} \right\rangle + x\intg [x]  \left\langle \frac{E_{4}^3}{\Delta^{12}} \right\rangle  $   \\ 
%\hline
&&\\[-0.5em]
\makecell{ $Z^{1}_{2,7} $} & -44264 & $  \frac{E_{4}^2 }{\Delta^{1845}} $   \\ 
 \hline
&&\\[-0.5em]
\makecell{ $Z^{2}_{2,7} $  } & -44148 & $ \frac{ 2E_{6} }{\Delta^{1840}} $  \\
 \hline
&&\\[-0.5em]
\makecell{ $Z^{3}_{2,7}$  } & -44032 & $ \frac{ E_{4} }{\Delta^{1835}} $  \\
 \hline
&&\\[-0.5em]
\makecell{ $Z^{4}_{2,7}  $ }  & -43916 & $ \frac{ 2E_{4}^2 E_{6}   }{\Delta^{1831}} $  \\ 
\hline
&&\\[-0.5em]
\makecell{ $Z^{1}_{2,9}  $ }  & -94192 & $ \frac{ E_{4}  }{\Delta^{3925}} $  \\ 
\hline
&&\\[-0.5em]
\makecell{ $Z^{2}_{2,9}  $ }  & -94076 & $ \frac{ 2E_{4}^2 E_{6}   }{\Delta^{3921}} $  \\ 
\hline
%&&\\[-0.5em]
%\makecell{ $Z^{3}_{2,9}  $ } & - & $    $  \\ 
%\hline
%&&\\[-0.5em]
%\makecell{ $Z^{13}_{2,5}  $ } & -1232 & $   \frac{ E_{4}^2 }{\Delta^{52}} $  \\ 
%\hline
%&&\\[-0.5em]
%\makecell{ $Z^{14}_{2,5} = X_{2,5} \#_{f} E(28)_K $ } & -1236 & $   \frac{ 2E_{6} }{\Delta^{52}} $  \\ 
%\hline
%&&\\[-0.5em]
%\makecell{ $Z^{15}_{2,5} $ } & -1240 & $   \frac{ E_{4} }{\Delta^{52}} $  \\ 
%\hline
\end{tabular}
\end{center}
\caption{Generators $E(X)$ of $\intg$-polynomial ring part of $\pi_{d}$TMF associated to $Z^{r}_{2,n}$ for the E-string theory (rank one) are listed. Four-manifolds used in the fiber summation $X_{2,n}$ have their generators as  $E(X_{2,3}) = 2E_{4}^2 E_6 / \Delta^{125},\, E(X_{2,5}) =  2E_{4} E_6 / \Delta^{620},\, E(X_{2,7}) =  2 E_6 / \Delta^{1747 },\, E(X_{2,9}) =  2E_6 / \Delta^{3765} $. } 
\end{table}

\subsection{Fiber sum manifold}

We move onto the fiber sum manifold $Z^{r}_{2,n}$. Using Appendix A and B, its TMF degree $d$ is easily calculated to be
$$
d = -\frac{4}{3} \left(97 n^3+2 n-87 r\right).
$$
We propose the following fiber sum formula. 
\begin{conjecture} Let $Z^{r}_{2,n}$ be the fiber sum manifold and $d=d_{r,n}$ be its TMF degree as above. For values of $r,n$ such that $d \notin 24\intg$ and the E-string theory compactified on $Z^{r}_{2,n}$, its fiber sum formula takes the following form.
\begin{gather*}
E( Z^{r}_{2,n} )  \circeq  \Delta^{ \frac{ \chi (  \Sigma_{f+b}  ) }{2} + s(n) + k(r) }    E( X_{2,n}) \ast E( E(2r)_K )\\
E_{4}^{p_1} \ast E_{4}^{p_2} = E_{4}^{p_1 + p_2 + c},\quad \text{if}\ p_1 + p_2 + c \geq 3,\, \text{use}\ E_{4}^3 = j\Delta\\
E_{6}^{w_1} \ast E_{6}^{w_2}  = E_{6}^{b},\quad w_1 + w_2 \equiv b\quad \text{mod}\, 2, \quad  (b=0,1)\\
\Delta^{m_1} \ast \Delta^{m_2}  = \Delta^{m_1 + m_2}\\
 c= \begin{cases}
1,\ n \in 6 \intg_{+} + 1\\
0,\ otherwise\\
\end{cases}\\
\begin{tabular}{ |c|c|c|c|c|c|c|c| } 
 \hline
$n$ & 3 &  5 & 7 & 9 & 11 & 13 & 15 \\
 \hline
$s(n)$ & 3 & 6 & 10 & 16 & 23 & 31 & 40 \\
 \hline
\end{tabular}\\
k = \begin{cases}
-1,\ r = \text{even}\\
0, \quad r = \text{odd}\\
\end{cases}\\
\end{gather*}
where $f$ and $b$ are genus of the fiber and base surfaces, respectively.
\end{conjecture}

\noindent \textbf{Remark.}  We note that for a fixed $n$, the overall factor is independent of specific value of $r$. It turns out that the power of the overall factor cannot be expressed as a linear combinations of the $r$ independent quantities $\chi (X_{2,n}), \sigma(X_{2,n}), \chi (\Sigma_{f+b})$ and $c_{1}^{2}(Z^{r}_{2,n})$ for both even and odd $r$.
\newline

\noindent The conjecture was checked using Table 8.\\

We illustrate the derivations behind Table 8. For example, we consider $Z^{2}_{2,3}$. After substituting its parameter values into the above formula, we get $d=-3268 \equiv 4$ mod 8. Applying the same method as in Section 4.1, $\pi_{-3268}$TMF $\cong \pi_{188}$TMF. It turns out $\pi_{188}$tmf is torsion free~\cite{DFHH}
$$
\pi_{188}\text{tmf}_{(2)} = \intg_{(2)}^8 \qquad \pi_{44}\text{tmf}_{(3)} = \intg_{(3)}^2
$$
where the periodicity of $\pi_{\ast}\text{tmf}_{(3)}$ is used. The prime 2 generator is $2E_4 E_6 \Delta^7$ and the prime 3 generator is $E_4 E_6 \Delta$. After localization, we get the generator displayed in Table 8.

%\clearpage

%$$
%X_{2,3} : \pi_{-212} TMF =  \intg[x] \left\langle \frac{2 E_{4} E_{6}}{\Delta^{10}} \right\rangle \qquad X_{2,5}  : \pi_{-1060} \text{TMF} = \intg/2\intg \oplus  \intg[x] \left\langle \frac{2 E_{4}E_{6}}{\Delta^{45}} \right\rangle
%$$
%
%$$
 %X_{3} : \pi_{-320} TMF = \intg / 3 \intg \left\langle \frac{\alpha^2 \beta}{\Delta^{14}} \right\rangle \oplus  \intg[x] \left\langle \frac{E_{4}^2}{\Delta^{14}} \right\rangle \qquad X_{5}  : \pi_{-1560} \text{TMF} = 24\intg +  x\intg[x] \left\langle \frac{24}{\Delta^{65}} \right\rangle
%$$

\section{TMF and $2d\ (0,1)$ theories}

Evidence for Conjecture 1.1 was found for the TMF degrees $d=0,1,\cdots ,15,24$ and $d= -1,-2,\cdots ,-20,-24$ in \cite{GPPV}. We fill the gaps between $15$ and $24$ and between $-20$ and $-24$. Furthermore, we extend the table, which is in Appendix C (Table 9). An important lesson from \cite{GPPV} is that, for $d>0$ many of 2d theories are realized as a sigma model whose target space is determined by generators of $\pi_{d}$TMF~\footnote{Generators for $d=0$ and $d=4$ have nontrivial denominators, which make their 2d theories not sigma models.} whereas this is no longer true for $d<0$. In this section, we describe how Table 9 is obtained for several examples.
\newline

For $d=16 \equiv 0$ mod 8, we see that $\pi_{16}\text{tmf}_{(2)}= \intg_{(2)}$ and $\pi_{16}\text{tmf}_{(3)}= \intg_{(3)}$ are generated by $E_{4}^2$~\cite{DFHH} (Chapter 13). Hence $\pi_{16}$ TMF $= \intg [x] \left\langle E_{4}^2 \right\rangle $. Using $\hat{A}(Y_8)=1 $ and the multiplicative property of the $\hat{A}$ genus under cartesian product of two manifolds, we see that $\hat{A}(Y_8 \times Y_8)=1 $ and its Witten genus of $Y_8 \times Y_8$  is $E_{4}^2$. Then the corresponding $2d\, (0,1)$ theory is a sigma model having target space~\footnote{Dimension of a target space is set by the degree $d$ of TMF.} as $Y_{8}^2$ (see Appendix C for the construction of $Y_8$).
\newline

For $d=17 \equiv 1$ mod 8, $\pi_{17}\text{tmf}_{(2)}= \intg / 2\intg \left\langle \eta E_{4}^2\right\rangle \oplus  \intg / 2\intg \left\langle \kappa\nu\right\rangle$ whereas prime 3 part is trivial. A string manifold corresponding to $\eta$ is $S^1$ and to $\nu$ is $S^3$; a Lie group $G_2$ is associated to $\kappa$~\cite{GPPV}. Consequently, we arrive at the sigma models written in Table 9.
\newline

For $d=20 \equiv 4$ mod 8, $\pi_{20}\text{tmf}_{(2)}= \intg / 8\intg \left\langle \bar{\kappa} \right\rangle \oplus  \intg_{(2)} \left\langle 2E_4 E_6\right\rangle$
whereas $\pi_{20}\text{tmf}_{(3)}= \intg / 3\intg \left\langle \beta^2 \right\rangle \oplus  \intg_{(3)} \left\langle E_4 E_6\right\rangle$. Then $\pi_{20}$ TMF $= \intg/ 24 \intg \oplus \intg [x]$. A Lie group $Sp(2)$ is associated with $\beta$~\cite{GPPV}. Hence we arrive at the target spaces in the table.
\newline

When $d= -23$, we need to find a theory corresponding to $8\eta /\Delta$ generator~\footnote{For $d<0$, a $2d\, (0,1)$ theory is not given by a sigma model~\cite{GPPV}} . Applying the strategy in \cite{GPPV}, we search for a 2d CFT with central charges $(23/2, 0)$ consisting of left moving fermions moving in $\real^r/ \Lambda$ in which $\Lambda$ is an odd self dual lattice. Specifically, this is may be realized by a chiral WZW theory at level $k \in \intg$ associated with a Lie algebra $g$ whose root lattice $\Lambda_R = \Lambda$. The central charge of the $g$ WZW theory is given by
$$
c = \frac{k\, \text{dim}\, g}{k+h^{\vee}}
$$ 
where $h^{\vee}$ is the dual Coxeter number. We find that $g=so(5)$ and $g=sp(4)$ both at $k=-23$. The fermions are moving in $\real^2/ \Lambda_2$. Since they are isomorphic, pick the former and denote it by $T_{(B_2 ;-23)}$ in Table 9 in Appendix C.
\newline

\indent We next consider a generator $\nu /\Delta^2$ whose $d= -45$. As done above, we find that WZW theories $T_{(B_3 ;-75)}$ and $T_{(B_{22} ;1)}$ have $c_L = 45/2$. They are in the same connected component and hence deformation equivalent. In the former theory, fermions are moving in $\real^3 / \Lambda_{3}$ whereas they are propagating in $\real^{22} / \Lambda_{22}$ in the latter.\\
%We move onto $d= -36$. For this case, we need to a 2d theory corresponding to $\alpha/\Delta$, which has $d=-21$. Possible WZW theories for this generator are $(g=B_2 ; k= -63), (B_3 ; 5), (B_{10} ; 1)$ and $(C_2 ; -63)$. They all have $c_L = 21/2$ and are all deformation equivalent. As above, we pick $(B_{10} ; 1)$ and arrive at $\pi_{-36}$TMF written in Table 8 in Appendix C.

\section{Future directions}	
 
 \begin{itemize}
	 \item For $Z^{r}_{g,n}$ manifold in which $g$ is odd, torsion generators of $\pi_{\ast}$tmf appear in the theories considered above. Fiber sum formulas in this case would take qualitatively different form including a multiplication rule.
	
	 \item Higher rank ($Q>1$) E-string theory possesses an extra symmetry and hence, its anomaly polynomial generalizes. The additional symmetry is expected to have nontrivial effects on fiber sum formula. 
	
	 \item Generally in order to find $2d\ (0,1)$ theories corresponding to $\pi_{\ast}$TMF for $\ast >24$, string manifolds whose Witten genus are $E_4 \Delta^{m}$ and $2E_6 \Delta^{n},\, m,n \in \intg_+ $ are needed. Constructions of the string manifolds are not known yet.
	
   %\item It would be interesting to find a fiber sum formulas for other smooth spin 4-manifolds.
	
 \end{itemize}

\textbf{Acknowledgments.} I am grateful to Sergei Gukov, Andre Henriques, Du Pei, Pavel Putrov, and Michael Hill for the numerous explanations. I would like to thank Sergei Gukov for reading a draft of this paper.   

\appendix
\section*{Appendix}
\addcontentsline{toc}{section}{Appendix}

\section{Four manifolds}

We review the fiber sum operation between 4-manifolds and summarize the topological data of the 4-manifolds considered in this paper.
\newline

A natural generalization of the connected sum operation for (smooth) surface bundles on a surface is the fiber sum~\cite{Gf}. For instance, it can be used to obtain new elliptic surfaces from given elliptic surfaces. Moreover, it was used to construct a family of mutually nondiffeomorphic 4-manifolds and to analyze the symplectic geography problem~\cite{AP, P, PS}. It appeared in disproving a conjecture (the minimal conjecture) regarding a classification of smooth closed 4-manifolds~\cite{Sz}. The fiber sum operation between two surface bundles on surfaces $X_1$ and $X_2$ is defined as
$$
X_1 \#_{f} X_2 = \lb X_1 \setminus \nu \Sigma_g \rb \cup_{\phi} \lb X_2 \setminus \nu \Sigma^{\prime}_g \rb
$$
where $\nu \Sigma_g$ and $\nu \Sigma^{\prime}_g$ are tubular neighborhoods of the surfaces and $\phi$ is an orientation reversing diffeomorphism.
\newline

\noindent \underline{Elliptic surfaces}\, An elliptic surface is a compact complex surface $S$ in which there is a holomorphic map to a complex curve $C$, $\pi : S \rarw C$  such that a generic fiber is a smooth elliptic curve~\cite{SS}. Elliptic surfaces have been classified up to deformation equivalence. A well-known example is $S=E(n)$ whose generic fiber is $T^2$. They are constructed inductively using the fiber sum operation starting from $E(1)$. 
$$
E(n) = E(n-1)  \#_{f} E(1).
$$
Topological data of E(n) are
$$
Q_{E(n)} = n (-E_8 ) \oplus (2n-1) H \qquad n= \text{even}
$$
$$
\chi (E(n)) = 12n \qquad \sigma (E(n)) = -8n \quad \text{for all}\ n
$$
$$
b_2 (E(n)) = 12n-2 \qquad b_{2}^{+} = 2n-1 \qquad b_1 = 0 \quad \text{for all}\ n
$$
Moreover, $E(n)$ are all simply connected. 
\newline

\noindent\underline{Further Surface bundles}\,  Using $E(n)$, the knot surgery and the fiber sum operation, infinite families of smooth simply-connected spin symplectic closed 4-manifolds that are mutually nondiffeomorphic were constructed in \cite{AP}. They are all irreducible (i.e. they cannot decomposed as a connected sum of two 4-manifolds in which neither of them are homotopy 4-spheres). We summarize the families.
$$
Z^{r}_{g,n} = X_{g,n} \#_{\Sigma_{f+b}} E(2r)_K \qquad g \geq 2 \quad n=\text{odd}\, \geq 3\\
$$
$$
r = \text{min} \lac \frac{1}{12} g(g-1)(n^2 -1) n^{2g-3}, 2+gn+g(g-1)(n-1)n^{2g-3} \rac
$$
where $f$ is a fiber genus, $b$ is a base genus and $X_{g,n}$ is
\begin{align*}
\Sigma_{gn} \rarw & X_{g,n} \\
& \downarrow \\
& \Sigma_{1+g(g-1)n^{2g-2}}
\end{align*}
Its signature $\sigma(X_{g,n})$ is $(4/3)g(g-1)(n^2 -1) n^{2g-3}$. $E(2r)_K$ is the Fintushel-Stern knot surgery~\cite{FS} on $E(2r)$ along a a fibered knot of a certain genus $g(K)$ in $S^3$. The signature and Euler characteristic of $Z^{r}_{g,n}$ are 
\begin{align*}
\sigma (Z^{r}_{g,n} ) & = \frac{4}{3}g(g-1)(n^2 -1)n^{2g-3} - 16r\\
e(Z^{r}_{g,n} ) & = 4gn(g(g-1)n^{2g-2} + 1 ) + 24r
\end{align*}
\newline

\noindent Another surface bundle analyzed in \cite{AP} was
$$
V^{r}_{n} = X_{n} \#_{\Sigma_{f+b}} E(2r)_K\qquad  n=\text{odd}\, \geq 3 
$$
$$
r = \text{min} \lac \frac{1}{6} (n+1)n(n-1), 2n^2 +n +2 \rac
$$
where
\begin{align*}
\Sigma_{3n} \rarw & X_{n} \\
& \downarrow \\
& \Sigma_{2n^2 +1}
\end{align*}
Its signature $\sigma(X_{n})$ is $(8/3)(n+1)n(n-1)$. The signature and Euler characteristic of $V^{r}_{n}$ are 
\begin{align*}
\sigma (V^{r}_{n} ) & = \frac{8}{3}(n+1)n(n-1) - 16r\\
e(V^{r}_{n} ) & = 12n(2n^2 +1) + 24r
\end{align*}
%
%\newline

\noindent \underline{Nonsymplectic manifolds}\, Smooth simply-connected closed oriented 4-manifolds that do not carry symplectic structures $Z(k,m)_n $ were constructed in \cite{Sz} $(k,m \in \intg_{+},\, n \geq 0)$. They have even intersection form for $k,m$ odd and hence carry spin structures $(w_2 = 0)$. The construction consists three steps. The first part is
\begin{align*}
Y & =  E(1) \#_{f} T^4  \#_{f} E(1) \\
  & =  \lb E(1) \setminus nd F_1 \rb \bigcup_{g_1} \lb T^4 \setminus (nd T_{12} \bigcup nd T_{13} ) \rb \bigcup_{g_2} \lb  E(1) \setminus nd F_2 \rb
\end{align*}
where $F_1$ and $F_2$ are generic fibers in $E(1)$'s, $T_{12}$ and $T_{13}$ are symplectic submanifolds, $g_1$ and $g_2$ are orientation reversing diffeomorphisms and $nd$ denotes an open tubular neighborhood. We again apply the fiber sum operation to $Y$.
\begin{align*}
Y(k,m)     & =   \lb E(k)  \setminus nd T_1 \rb \bigcup_{g_1} \lb Y \setminus (nd L_1 \bigcup nd L_2 ) \rb \bigcup_{g_2} \lb  E(m) \setminus nd T_2 \rb
\end{align*}
The final step is to use the log. transform on $Y$, we obtain 
\begin{align*}
Z(k,m)_n & =   \lb E(k)  \setminus nd T_1 \rb \bigcup_{g_1} \lb X_n \setminus (nd L_1 \bigcup nd L_2 ) \rb \bigcup_{g_2} \lb  E(m) \setminus nd T_2 \rb
\end{align*}
where $X_n = \lb Y \setminus nd T_{14} \rb \bigcup_{\phi_n} \lb T^2 \times D^2 \rb$. Topological invariants of $Z(k,m)_n$ are 
$$
\chi_h (Z(k,m)_n ) =k+m+4 \qquad \sigma (Z(k,m)_n ) = -8(k+m) - 16.
$$

\section{Anomaly polynomials}

We summarize the anomaly polynomials described in \cite{GPPV} (see also \cite{CDI, RSSZ}). They provide 2-dimensional gravitational anomaly $2(c_L - c_R)$ expression in terms of compact 4-manifolds' topological invariants for $2d\ (0,1)$ theories that are obtained from compactification of $6D\ (1,0)$ SCFTs on the 4-manifolds. Perturbative anomaly of $6D\ (1,0)$ SCFTs can be characterized by polynomials in characteristic classes of various bundles on $6D$ spacetime manifold $M$.  Their anomaly polynomials are differential 8-form $I_8$. Its formula in our context is
$$
I_8 = \alpha c_2 (R)^2 + \beta c_2 (R) p_1 (T) + \gamma p_1 (T)^2 + \delta p_2 (T)  + I^{\text{flavor}},\quad \alpha,\beta,\delta,\gamma \in \real
$$
The second Chern class $c_2 (R) \in H^4 (R) $ is associated with the R-symmetry bundle $R$. Pontryagin classes of tangent bundle $TM$, $p_1 (T) \in H^4 (T)$ and $p_2 (T) \in H^8 (T)$ correspond to gravitational anomaly~\cite{GW}. $c_2 (R) p_1 (T)$ represents a mixed anomaly between R-symmetry transformation and diffeomorphism of $M$. The last term contains a flavor symmetry anomaly. 
\newline

Compactifications to 2-dimensions result in an anomaly polynomial that is a 4-form $I_4$. When a background field associated with flavor symmetry group is turned off, 2d gravitational anomaly is given by
$$
c_R - c_L = 18(\beta - 8 \gamma - 4\delta)\sigma(X) + 12\beta \chi (X)
$$
where $\sigma(X)$ is the signature of a 4-manifold $X$,  $\chi (X)$ is the Euler characteristic of $X$.

\noindent For the three theories analyzed in this paper, their anomaly polynomials are~\cite{CDI, OST, OSTY, GPPV}

\begin{itemize}
	\item Hypermultiplet $I_8 = \frac{7 p_1 (T)^2 - 4 p_2(T) }{5760} $
	
	\item Vector multiplet $I_8 = -\frac{1}{24}c_{2}(R)^2 - \frac{1}{48}c_{2}(R) p_1 (T)  - \frac{7 p_1 (T)^2 - 4 p_2(T) }{5760} $
	
	\item Tensor multiplet $I_8 = \frac{1}{24}c_{2}(R)^2 + \frac{1}{48}c_{2}(R) p_1 (T) + \frac{23 p_1 (T)^2 - 116 p_2(T) }{5760} $
	
	\item E-strings (rank $Q=1$) $I_8 = \frac{13}{24} c_2(R)^2 -  \frac{11}{48} c_2 (R) p_1 (T) + \frac{29(7p_1 (T)^2 - p_2 (T))}{5760} $
	
\end{itemize}
The fourth expression excludes the contribution of the free hypermultiplet.

\section{Topological modular forms}

We summarize properties of topological modular forms (see \cite{DFHH, H, Hen, G, L, GPPV} for details). There are three kinds of topological modular forms, tmf,  Tmf, and TMF. The first and the third variants are relevant for this paper. 

\begin{enumerate}

	\item tmf
	
	\begin{itemize}
	
		\item Connective property $\pi_{\ast <0} \text{tmf} = 0$
		
		\item Connective cover of Tmf. 
		
		\item Relation to the stable homotopy groups of spheres and the ring of modular forms 
		$$
		\pi^{s}_{\ast}\ \mathbb{S} \rarw \pi_{\ast} \text{tmf}  \rarw \text{MF}_{\frac{\ast}{2}}
		$$ 
		The first ring homomorphism is an isomorphism for $n = 0,1,\cdots ,6$. The second map is the boundary homomorphism/the elliptic genus. (it maps to zero for $\ast$ odd, thus the kernel is the torsion part).
		
		\item Relation with string manifolds $\Omega^{\text{string}}_{\ast} \twoheadrightarrow \pi_{\ast} \text{tmf} $ : every tmf classes comes from  a bordism class of string manifold (this surjective map is the topological Witten genus)\footnote{More precisely, this map $\pi_{\ast} MString \rarw \pi_{\ast} tmf$ is a surjective homomorphism}.  $\Omega^{\text{string}}_{\ast}$ is a $\ast$-dimensional string (co)bordism group. 
		
		\item Witten genus of a string manifold $\phi_W (Y)$ is $$\pi_{\ast} \text{MString} \rarw \pi_{\ast} \text{tmf}   \rarw \text{MF}_{\frac{\ast}{2}}$$ It is factored by the elliptic genus (MString is the Thom spectrum).
		
		\end{itemize}
	
	\item TMF
	 
	 \begin{itemize}
	
		 \item Localization $\pi_{\ast} \text{TMF} \simeq \pi_{\ast}  \text{tmf} \lsb \Delta^{-24} \rsb $ 
		
		\item Periodicity $\pi_{\ast } \text{TMF} \simeq  \pi_{\ast + 576} \text{TMF}$ ($ \Rightarrow \pi_{\ast < 0 } \text{TMF}$ can be nontrivial).
		
		\item $\pi_{n } \text{TMF} \supseteq \intg [x]$ for $n \equiv 0,4$\, mod $8$,\quad $\pi_{n } \text{TMF} \supseteq (\intg /2\intg) [x]$ for $n \equiv 1,2$\, mod $8$.
	
	 \end{itemize}

\end{enumerate}
%\newline
\noindent \underline{Relations between generators} 
$$
\pi_{1}(\mathbb{S}) = \intg / 2\intg \left\langle \eta \right\rangle \qquad \pi_{3}(\mathbb{S}) = \intg / 24\intg \left\langle \nu \right\rangle
$$
$$
\eta^3 = 12\nu \qquad 2\eta = 24\nu = 2\nu^2 = \eta^4 = \nu^4 =0
$$
$\eta$ is associated to the $S^3$-Hopf fibration, $S^1 \rarw S^3 \rarw S^2$ and $\nu$ corresponds to the $S^7$-Hopf fibration, $S^3 \rarw S^7 \rarw S^4$.
\newline

\noindent \underline{String manifold} We briefly review string manifolds (see \cite{DFHH, DHH} for details). A manifold equipped with a string structure is called a string manifold $Y$. Equivalent characterizations of the structure are 

\begin{itemize}

\item Lifting of the map $\tau$ to the classifying space $BO(n)$ : 

\begin{tikzcd}
& BString(n) \ar[d] \\
Y \ar[r, "\tau"'] \ar[ur]
& BO(n)
\end{tikzcd}
\newline
where $String(n)$ is the 6-connected cover of $O(n)$.

\item An extension of a spin structure of $Y$: $\frac{p_1 (TY)}{2} = 0 (w_1  = w_2 = 0)$.
	
\end{itemize}
\noindent The string condition is an orientablility condition for a generalized cohomology theory, which is TMF in our context.
\newline

%\noindent We list string manifolds and Lie groups associated with generators of $\pi_{\ast}$TMF that appear in this paper and/or in \cite{GPPV} (see \cite{DFHH} Chapter 13 for details of the generators). For torsion part of $\pi_{\ast}$TMF,  Lie group are associated. String manifolds (denoted by $Y_d$) are tied to torsion-free generators.
%\newline

%\begin{table}[t]
%\begin{center}
%\begin{tabular}{ |c|c| } 
 %\hline
%TMF generators & Manifolds    \\ 
 %\hline
%$ \eta $ & $ U(1) $  \\ 
 %\hline
%$ \nu, \alpha $ & $ SU(2) $  \\ 
 %\hline
%$ E_4 $ & $ Y_8 $  \\ 
 %\hline
%$ 2E_6 $ & $ Y_{12} $  \\ 
 %\hline
%$ 24\Delta $ & $ Y_{24} $  \\ 
 %\hline
%$ \epsilon $ & $ SU(3) $  \\ 
 %\hline
%$ \beta $ & $ Sp(2) $  \\ 
 %\hline
%$ \kappa $ & $ G_2 $  \\ 
 %\hline
%$ \overline{\kappa} $ & $ Sp(2) \times Sp(2) $ \\ 
 %\hline
%$ q $ &  $SU(3) \times SU(5)$ \\ 
 %\hline
%\end{tabular}
%\end{center}
%\end{table}
%\newline

\noindent \underline{Construction}\ We summarize on the construction of $Y_8$ described in \cite{GPPV}\footnote{$Y_{12}$ is constructed similarly.}.
	
	\begin{enumerate}
	
		\item Plumbing and parallelizable manifold: place a twisted $D^4$-bundle $E^8$ on $D^4$ at each vertex of the $-E_8$-graph. Applying the plumbing along each edge yields a 8-dimensional smooth manifold $W^8$ with an exotic $S^7$-boundary~\cite{M}. Since the $S^7$ represents an element of a subgroup $bP_{8} \subset \Theta_7$, $W$ is parallelizable~\cite{KM,M2}.
		
		\item Boundary sum and gluing: the cobordism group of homotopy 7-spheres is $\Theta_7 = \intg /28\intg$~\cite{KM}, we take a boundary sum of 28 copies of $W^8$ and then glue an $D^8$ along the $S^7$ to obtain $Y_8 = \natural^{28} W \cup D^8.$  The signature of $Y_8$ is $\sigma = -8\times 28$.
		
		\item Pontryagin number:  $W$ being parallelizable implies that $Y_8$ is almost parallelizable, which in turn means that only the top Pontryagin class $p_2$ of $Y_8$ is nonzero~\cite{KM2}. Hence, $Y_8$ carries a string structure (i.e. $p_1 (Y_8)/2=0$).  By the Hirzebruch signature theorem $\sigma(Y_8) = \left\langle L (p_1, p_2) , [Y_8] \right\rangle$,\,  $p_2 = -1440$ (only $L_2$ contributes).
		
		\item $\hat{A}$-genus and the Witten genus $\phi_W$: The above information enables to find $\hat{A}_2 = \frac{1}{5760} ( -4p_2 + p_{1}^2) = 1$, which agrees with the leading term of $\phi_W (Y_8) = E_4$.

	\end{enumerate}

%\item $Y_{12}$: We consider a twisted $D^6$-bundle $E^{12}$ on $D^6$ at each vertex of the $-E_8$-graph. After plumbing, we get a 12-dimensional smooth manifold $W^{12}$ with an exotic $S^{11}$-boundary~\cite{M}. Since the cobordism group of homotopy 11-spheres is $\Theta_{11} = \intg /992 \intg$~\cite{M2}, we take a boundary sum of 992copies of $W^{12}$ and then glue an $B^{12}$ along the $S^{11}$ to obtain $Y_{12}$:
%$$
%Y_{12} = \natural^{992} W^{12} \cup B^{12}. 
%$$

%\item $Y_{24}$: constructed homotopically in \cite{MH} (Chapter 9).

\begin{table}[h!]
\begin{center}
\begin{tabular}{ |c|c|c| } 
 \hline
$d$ & $ \pi_d \text{TMF}$  & $2d\ \mathcal{N}=(0,1)$ theory generators  \\ 
 \hline
&&\\[-0.5em]
$16$  &  $ \intg[x] $  & $\sigma (Y_8 \times Y_8 ) $\\
 \hline
&&\\[-0.5em]
$17$  & $ \intg /2 \intg \oplus (\intg /2 \intg) [x] $  & $\sigma ( SU(2) \times G_2 ),\, \sigma ( S^1 \times Y_8 \times Y_8 )$ \\
 \hline
&&\\[-0.5em]
$18$  & $ \intg /2 \intg [x]  $   & $ \sigma ( T^2 \times Y_8 \times Y_8 ) $ \\
 \hline
&&\\[-0.5em]
$19$  & $ 0 $  &  - \\
 \hline
&&\\[-0.5em]
$20$  &  $ \intg/ 24\intg  \oplus \intg[x] $ &  $ \sigma (Sp(2) \times Sp(2)) ,\,  \sigma (Y_8 \times Y_{12}) $ \\
 \hline
&&\\[-0.5em]
$21$  &  $ \intg /2 \intg $ & $ \sigma (Sp(2) \times Sp(2) \times U(1) ) $ \\
 \hline
&&\\[-0.5em]
$22$  & $ \intg /2 \intg $   &  $ \sigma ( SU(3) \times G_2 ) $ \\
\hline
&&\\[-0.5em]
$23$  & $ 0 $ & - \\
\hline
&&\\[-0.5em]
$28$  &$ \intg /2 \intg \oplus  \intg [x] $   & $ \sigma ( G_2 \times G_2 ) ,\, \sigma ( Y_{8} \times Y_{8} \times Y_{12} )  $ \\
\hline
&&\\[-0.5em]
$29$  & $ 0 $   & -  \\
\hline
&&\\[-0.5em]
$30$  &$ \intg / 3 \intg  $   & $ \sigma ( Sp(2) \times Sp(2) \times Sp(2) ) $  \\
\hline
&&\\[-0.5em]
$31$  & $ 0 $ & - \\
\hline
&&\\[-0.5em]
$34$  & $ \intg / 2\intg \oplus (\intg /2 \intg) [x]  $ & $\sigma ( Sp(2) \times Sp(2) \times G_2 ),\,  \sigma ( T^2 \times Y_{32}) $ \\
\hline
&&\\[-0.5em]
$-21$  & $ 0 $ & - \\
\hline
&&\\[-0.5em]
$-22$  &  $ (\intg /2 \intg) [x] $  & $\sigma (U(1)) \otimes	T_{(B_2;-23)}$  \\
\hline
&&\\[-0.5em]
$-23$  &  $(\intg /2 \intg) [x]$  &  $T_{(B_2;-23)}$  \\
\hline
&&\\[-0.5em]
$-45$  &  $\intg /24 \intg$  &  $T_{(B_{22};1)}$  \\
\hline
\end{tabular}
\caption{This table lists $\pi_{d}$TMF and corresponding $2d\ (0,1)$ supersymmetric QFTs. It fills the gaps in the table in \cite{GPPV} and extend it. Although construction of $Y_{32}$ is yet unknown, it is an element of the string bordism group $\Omega^{\text{string}}_{\ast}$ by the surjective property of the map to $\pi_{\ast}$tmf (see the review above). The second column can be obtained using \cite{DFHH} (Chapter 13).}
\end{center}
\end{table}

\end{document}